\documentclass[a4paper]{article}

\usepackage[margin=1in]{geometry}
\usepackage{amsfonts}
\usepackage{amssymb,amsmath}

\usepackage[T1]{fontenc}
\newcommand{\changefont}[3]{
\fontfamily{#1} \fontseries{#2} \fontshape{#3} \selectfont}

\changefont{ptm}{m}{n}

\usepackage{setspace} 
 \usepackage{graphicx}    % needed for including graphics e.g. EPS, PS

\newcommand \be{\begin{equation}}
\newcommand \ee{\end{equation}}
\newcommand \ba{\begin{eqnarray}}
\newcommand \ea{\end{eqnarray}}

\def\bit{\begin{itemize}}
\def\eit{\end{itemize}}

\usepackage{amsfonts}
\usepackage{amssymb}
\usepackage{graphics}
\usepackage{mathrsfs}
\usepackage{color}

\newtheorem{theorem}{Theorem}[section]

\long\def\symbolfootnote[#1]#2{\begingroup%
\def\thefootnote{\fnsymbol{footnote}}\footnote[#1]{#2}\endgroup} 

\begin{document}

\begin{center}
\Large \textbf{Exogenous Versus Endogenous for Chaotic Business Cycles}
\end{center}

\vspace{-0.3cm}
\begin{center}
\normalsize \textbf{Marat Akhmet$^{a,} \symbolfootnote[1]{Corresponding Author Tel.: +90 312 210 5355, Fax: +90 312 210 2972, E-mail: marat@metu.edu.tr}$, Zhanar Akhmetova$^b$, Mehmet Onur Fen$^c$} \\
\vspace{0.2cm}
\textit{\textbf{\footnotesize$^a$Department of Mathematics, Middle East Technical University, 06800, Ankara, Turkey}} \\
\textit{\textbf{\footnotesize$^b$Department of Economics, Australian School of Business, University of New South Wales, Sydney, NSW 2052, Australia}} \\
\textit{\textbf{\footnotesize$^c$Neuroscience Institute, Georgia State University, Atlanta, Georgia 30303, USA}}
\vspace{0.1cm}
\end{center}

\vspace{0.3cm}

\begin{center}
\textbf{Abstract}
\end{center}

\noindent\ignorespaces

We propose a novel approach to generate chaotic business cycles in a deterministic setting. Rather than producing chaos endogenously, we  consider aggregate economic models with limit cycles and equilibriums, subject them to chaotic exogenous shocks and obtain chaotic cyclical motions. Thus, we emphasize that chaotic cycles, which are inevitable in  economics, are  not  only interior properties of economic models, but also can be considered as a result of interaction of several economical systems. This provides a comprehension of chaos (unpredictability, lack of forecasting) and control of chaos as a global economic phenomenon from the deterministic point of view.
 
We suppose that the results of our paper are contribution to the mixed exogenous-endogenous theories of business cycles in classification by P.A. Samuelson \cite{Samuelson}. Moreover, they demonstrate that the irregularity of the extended chaos can be structured, and this distinguishes them from the generalized synchronization. The  advantage  of the knowledge of the structure is that by applying instruments, which already have been developed for deterministic chaos one can control the chaos, emphasizing a parameter or a type of motion. For the globalization of cyclic chaos phenomenon we utilize new   mechanisms such that entrainment by chaos, attraction of chaotic cycles by equilibriums and bifurcation of chaotic cycles developed in our earlier  papers.

\vspace{0.2cm}
 
\noindent\ignorespaces \textbf{Keywords:} Business cycle models; Exogenous shocks; Period-doubling cascade; Attraction of chaotic cycles; Chaotic business cycle

\section{Introduction}

Business cycles are a commonly accepted phenomenon in economics. However, we do not actually observe perfectly periodic motions in economic variables. Instead, economic data is highly irregular. One way to reflect this in economic models is to allow for stochastic processes. Deterministic differential equations can also be turned into a better picture of economic reality by introducing chaos.\footnote{There exists a third approach, which is somewhere in-between the two, where iterated function systems generated by the optimal policy functions for a class of stochastic growth models converge to invariant distributions with support over fractal sets \cite{Mitra09}.}

Chaotic economic systems can be viewed as unpredictable due to their sensitivity to initial values, which makes forecasting extremely difficult \cite{Baumol,Brock,Goodwin90,Rosser}. This is known also as the {\it butterfly effect} \cite{Lorenz63}. Devaney \cite{Dev90} proposed that {\it sensitivity} in conjunction with other properties, namely {\it transitivity} and {\it density of periodic solutions}, be considered as ingredients of chaos. An alternative way to prove the presence of chaos is by observing the {\it period-doubling cascade} \cite{Gleick87}. This chaos is also sensitive, since there are  infinitely many solutions with different periods and they are {\it unstable}. We utilize these ways of observing chaos in our paper. Importantly, irregularity based on theoretical {\it deterministic chaos} can be visualized in simulations. 

One should remark that it is not only sensitivity that can be considered as a mathematical representation of unpredictability, but  also the existence of infinitely many {\it unstable} periodic solutions. Indeed, while the presence of a single periodic solution can be accepted as a strong indicator of predictability (if one knows the values of the process during the period, then one knows all its future values), with infinitely  many {\it unstable} periodic solutions all the cycles are unstable, and the trajectory  of the dynamics wanders around, visiting  neighborhoods of the cycles in an unpredictable way. That is the reason why in the literature the proof of the existence of a periodic-doubling cascade is accepted as evidence of chaos. Stabilizing  periodic solutions is named in chaos theory as control of chaos.

Chaos theory could provide a new approach to economic policy-making. Economists believed initially that chaotic dynamics is not only unpredictable, but also un-controllable. The results of  Ott et al. \cite{Ott90} showed that  control of a chaos can be made by  a very  small corrections of parameters \cite{Gon04,holyst1}. This and related methods have been widely applied to economic models, as exemplified by Holyst et al. \cite{holyst}, Kaas \cite{Kaas}, Mendes and Mendes \cite{Mendes05}, Chen and Chen \cite{Chen07} and many others.

In the classic book \cite{Samuelson} it is observed that while forced oscillator systems naturally emerge in theoretical investigations of several technical and physical devices, economic examples for this special family  of functions have only  rarely been provided. The main reason for this deficiency may lie in the fact that the necessary  periodicity of the dynamic forcing may not be obvious in most economic applications. Our proposals are to apply \textit{deterministic and chaotic} exogenous shocks to economic models and make them more realistic.

One may view chaos (the lack of forecasting) as undesirable in economics, but unavoidable. Hence a deterministic economic model is realistic if it exhibits chaotic motions. We suggest considering the presence of chaos in a model not only as an indication of its adequacy, but also as a measure of its power. Indeed, the presence of chaos implies that the model generates infinitely many aperiodic motions and motions with different periods, which are unstable, and consequently easily affected by control and sustained in a desirable mode. In other words, deterministic chaos is essential for the flexibility and high-speed adjustment of economic models, an indispensable feature in the modern world.

The principal novelty of our investigation is that we create a chaotic perturbation, plug it in a regular dynamic system, and find that similar chaos is inherited by the solutions of the new system. We call this as \textit{the input-output mechanism of chaos generation}. This approach has been widely applied to differential equations before, but for regular inputs. In the studies \cite{Akh5,Akh2,Akh1,Akh7,Akh6}, the mechanisms for generating chaos in systems with asymptotically stable equilibria are provided. In contrast, in \cite{Kydland,Liu,Long,Mircea} unpredictability in the solutions of differential equations was considered a result of random perturbations with small probability.

P.A. Samuelson \cite{Samuelson} accepts purely endogenous theory as ``self-generating'' cycle. Following this opinion we understand chaos as endogenous if it is self-generated by an economic model. One can find detailed analysis of the endogenous chaos in  books \cite{lorenz,Rosser,zhang} and paper \cite{Baumol}, which are very seminal sources  on the  subject. The dynamics arise  in  duopoly models \cite{Rand}, in simple  ad hoc macroeconomic models \cite{DayShafer83,Stutzer}. By applying the Li-Yorke theorem it is shown in \cite{Benhabib80, Benhabib82} that an overlapping generations model of the Gale type could generate endogenous chaotic cycles. Discrete equations have been applied to investigate the presence of chaos in papers \cite{Day82,Day83}, where models representing a capital stock with a maximum capital-labor  ratio and a Malthusian  agrarian economy are investigated. In \cite{Dana,Day83,Pohjola} endogenous chaotic cycles are demonstrated in growth cycle models. The multiplier-accelerator model of Samuelson \cite{Samuelson} has been modified for generation of chaotic endogenous cycles and investigated in \cite{Blatt, Gabish84, Nusse}. Investigations in Kaldor's type models, which are originated from \cite{Hicks,Blatt} and finalized in \cite{Brock88}, showed that they could generate endogenous chaos.

Economists of the first half of the last century already felt a strong need for a theory of irregularities, particularly of irregular business cycles. In his classic book, Samuelson \cite{Samuelson} observes that while forced oscillator systems naturally emerge in theoretical investigations of several technical and physical devices and phenomena, economic examples for this special family of functions have only rarely been provided. The main reason for this dearth of evidence may lie in the fact that the necessary periodicity of the dynamic forcing may not be obvious in most economic applications. That is, economic phenomena do not display the kind of regularity that physical phenomena do. Samuelson \cite{Samuelson} states that ``... in a physical system there are grand conservation laws of nature, which guarantee that the system must fall on the thin line between stability and instability. But there is nothing in the economic world corresponding to these laws ...". In a passage Samuelson \cite{Samuelson} suggests that ``It is to be stressed that the exogenous impulses which keep the cycle alive need not themselves be even quasi-oscillatory in character." Thus, he was already talking about irregular business cycles that emerge as a result of \textit{irregular exogenous shocks}. Moreover, he recognized that ``most economists are eclectic and prefer a combination of endogenous and exogenous theories." Accordingly, in the present paper we consider economic models that admit \textit{endogenous} business cycles and are perturbed by \textit{exogenous} chaotic disturbances. Examples of models possessing limit cycles are Kaldor-Kalecki models and Lienard type equations with relaxation oscillations which are popular in economics. Next, the systems are subject to \textit{exogenous} chaotic disturbances, sensitive and with infinitely many unstable periodic solutions.

We propose two techniques of obtaining exogenous chaotic cycles as solutions of differential equations. In the first approach, an economic model with a limit cycle is perturbed chaotically to produce a chaotic business cycle. In the second one, we consider a system with an equilibrium, perturb it by cyclic chaos and observe that a chaotic business cycle emerges as a result. While the first method is theoretically verified in \cite{Akh8}, the second method of cyclic chaos generation is new and is demonstrated in our paper through simulations. Currently, we study cases where the shocks enter the system additively, but future investigations may involve more complex scenarios, where the disturbance enters the main functions of the economic model.

Goodwin \cite{Goodwin90} argues that the apparent unpredictability of economic systems is due to deterministic chaos as much as to exogenous shocks.  In this  sense, our  results can  be  interpreted  as  the  {\it  transmission of unpredictability}  from one  economic system to another, and  even models that  do not admit   irregularity in isolation can  eventually  be contaminated with  chaos.  Thus,  we provide support to the idea that unpredictability is  a {\it global  phenomenon} in  economics, and demonstrate one of the mechanisms  for  this  contagion. Considering the current extensive globalisation process, this is a good depiction of reality.

Our results demonstrate that the control may become not a local (applied to an isolated model) but a global phenomenon with strong effectiveness such that control applied to a model, which is realizable easily (for example, the logistic map or Feichtinger's generic model), can be sufficient to rule the process in all models joined with the controlled one. Another benefit of our studies is that in literature controls are applied to those systems, which are simple and low-dimensional. Control of chaos becomes difficult as the dimensions of the systems increase and the construction of Poincar\'{e} sections becomes complicated. Chaos control cannot be achieved if we do not know the period of unstable motion to be controlled.  In our  case, the control is applicable to models of arbitrary dimensions as long as the basic period of the generator is known. For these reasons, the possibility to control generated chaos by controlling the exogenous shocks that produce the said chaos is appealing. It is especially appealing from a policy-maker's point of view, as it offers a cost-effective way to regulate an economic system.

Control of chaos is nowadays  a synonym to the suppression of chaos. Thus our results give another way of suppression of chaos. If we find the controllable link (member) in a chain (collection) of connected chaotic systems, then we can suppress chaos in the whole chain. This is the effective consequence of our studies.

\subsection{Organization of the paper}

The paper is organized in the following way. In the next section we describe the input-output mechanism that serves as the basis of chaos extension and formulate two theorems that provide theoretical support to the subsequent discussion. In Section \ref{econ_sec3} economic models with regular motions - stable equilibrium and orbitally stable cycle - are introduced. These models are chaotically perturbed in the following section to obtain the main economic dynamics of the paper. More precisely, Section \ref{econ_sec4} considers a constellation of five economic models connected unilaterally. The extension of chaos near an equilibrium attractor, the entrainment by chaos of limit business cycles, the bifurcation of a chaotic cycle, and the attraction of a chaotic cycle are the scenarios of the appearance of chaos, and in some cases of chaotic business cycles, in economic models that we demonstrate. The effects of applying OGY control \cite{Ott90} to the models will also be presented. Section \ref{Kaldor_Kalecki_delay} provides simulation results for the entrainment by chaos of limit cycles of economic models with time delay. We compare in detail our method of chaos generation with that based on the synchronization of chaos \cite{Kocarev96,Pecora90,Rulkov95} in Section \ref{econ_sec6}. In particular, we argue that chaotic business cycles in the paper cannot be obtained by the synchronization of chaotic systems. In Section \ref{econ_sec7} we discuss our results from the point of view of self-organization, and particularly synergetics of Haken \cite{haken83}. We summarize the obtained results in the Conclusion.

\section{The Input-Output Mechanism and Its Applications} \label{econ_sec2}
 
To explain the input-output  mechanism  of chaos generation, let us  introduce   systems, which  we  call {\it the  base-system,  the  replicator  and  the  generator}.   They    are intensively  used in the manuscript. Consider the following  system of differential  equations, 
\begin{eqnarray} 
\label{base-system}
z' = B(z),
\end{eqnarray}
where $B:\mathbb R^n \to \mathbb R^n$ is a continuously differentiable function. The system  (\ref{base-system}) is called {\it the base-system}. 

Next,  we  subdue the   base-system  to a perturbation, $I(t),$ which will be called an {\it input}  and obtain the  following system,
\begin{eqnarray} 
y' = B(y) + I(t),\label{intro_replicator}
\end{eqnarray}
 which  will  be called as   {\it the replicator}.

Suppose that the input $I(t)$ admits a certain property, say, it is a bounded function. Assume that there exists a unique solution, $y(t),$ of the replicator system (\ref{intro_replicator}), with the same property. This solution is called an \textit{output}. The process of obtaining the solution $y(t)$ by applying the perturbation $I(t)$ to the base-system (\ref{base-system}) is called \textit{the input-output mechanism}. It is known that for certain base-systems, if the input is a periodic, almost periodic, bounded function, then there exists an output that is also a periodic, almost periodic, bounded function. In our paper, we consider inputs of a different nature: chaotic functions and set of cyclic chaotic functions. The motions that are in the chaotic attractor of the Lorenz system \cite{Lorenz63}, considered altogether, give us an example of a chaotic set of functions. Each element of this set is considered as a chaotic function. Both a set of functions and a single function can serve as an input (as well as an output), and we will use both types of inputs and outputs in this study.

We consider base-systems of two kinds: (i) systems with asymptotically stable equilibria, (ii) systems with limit cycles. In the former case, we will talk about attraction of chaos by equilibria, and in particular, attraction of cyclic chaos by equilibria. If the base-system admits a limit cycle, then we talk about the entrainment by chaos of limit cycles or just about entrainment by chaos \cite{Akh8}. If the limit cycle in a base-system is the result of a Hopf bifurcation \cite{Hassard}, we will also talk about the bifurcation of the cyclic chaos.

In our previous papers \cite{Akh5,Akh9,Akh7,Akhmet2012,Akh6} we analyzed the extension of chaos when the base system possesses an asymptotically stable equilibrium. The present paper focuses mostly on the generation of cyclic chaos through unilateral coupling of multiple systems.

The main source of chaos in theory are difference and differential equations. For this reason we consider in our manuscript, inputs, which are solutions of some systems of differential or discrete  equations equations. These systems will be called {\it generators}.\footnote{In future work, economic time series that have been tested for the presence of deterministic chaos may be considered (see \cite{Brock,Brock96,Decoster92,Panas,Wei98}.)}  

Thus, we consider the following system of differential equations, 
\begin{eqnarray} \label{intro_generator}
x' =  G(t,x),
\end{eqnarray}
where the function $G: [0, \infty) \times \mathbb R^m \to \mathbb R^m$ is continuous in all of its arguments. 
We assume that system (\ref{intro_generator}) possesses a chaotic attractor, and we call this system a {\it generator}. If $x(t)$ is a solution of the system from the chaotic attractor, then we take $$I(t) = \varepsilon \psi(x(t)),$$ and use the function $I(t)$ in equation (\ref{intro_replicator}). Here, $\epsilon$ is a non-zero real number and the function $\psi: \mathbb R^m \to \mathbb R^n$ is continuous. Since we use $x(t)$ as a perturbation in the network (\ref{intro_replicator}), we call it a \textit{chaotic solution}. The chaotic solutions may be irregular as well as regular (periodic and unstable) \cite{Dev90,Feigenbaum80,Sander11,Sander12}. In this study we will utilize also the logistic map \cite{Dev90} as a generator.

System (\ref{intro_generator}) is called sensitive if there exist positive numbers $\epsilon_0$ and $\Delta$ such that for an arbitrary positive number $\delta_0$ and for each chaotic solution $x(t)$ of (\ref{intro_generator}), there exist a chaotic solution $\overline{x}(t)$ of the same system and an interval $J \subset [0,\infty),$ with a length no less than $\Delta,$ such that $\left\|x(0)-\overline{x}(0)\right\|<\delta_0$ and $\left\|x(t)-\overline{x}(t)\right\| > \epsilon_0$ for all $t \in J.$

For a given chaotic solution $x(t)$ of (\ref{intro_generator}), let us denote by $\eta_{x(t)}(t,y_0),$ $y_0\in\mathbb R^n,$ the solution of (\ref{intro_replicator}) with $\eta_{x(t)}(0,y_0)=y_0.$ System (\ref{intro_replicator}) replicates the sensitivity of (\ref{intro_generator}) if there exist positive numbers $\epsilon_1$ and $\overline{\Delta}$ such that for an arbitrary positive number $\delta_1$ and for each solution $\eta_{x(t)}(t,y_0),$ there exist an interval $J^1\subset [0,\infty),$ with a length no less than $\overline{\Delta},$ and a solution $\eta_{\overline{x}(t)}(t,y_1)$ such that $\left\|y_0-y_1\right\|<\delta_1$ and $\left\|\eta_{x(t)}(t,y_0)-\eta_{\overline{x}(t)}(t,y_1)\right\| > \epsilon_1$ for all $t \in J^1.$ Moreover, we say that system (\ref{intro_replicator}) is chaotic if it replicates the sensitivity of (\ref{intro_generator}) and the coupled system $(\ref{intro_generator})+(\ref{intro_replicator})$ possesses infinitely many unstable periodic solutions in a bounded region.   

Next, we will formulate a theorem that forms the mathematical basis of the paper.

The following conditions are required:
\begin{enumerate}

\item[\bf (C1)] System (\ref{base-system}) admits a non-constant and {\it orbitally stable} periodic solution;
\item[\bf (C2)] System (\ref{intro_generator}) possesses sensitivity and is chaotic through period-doubling cascade;
\item[\bf (C3)] The functions $B$ and $G$ are bounded;
\item[\bf (C4)] There exists a positive number $L_B$ such that 
\begin{displaymath}
\left\|B(z_1)-B(z_2)\right\| \leq L_B\left\|z_1-z_2\right\|,
\end{displaymath}  
for all $z_1,$ $z_2 \in \mathbb R^{n};$ 
\item[\bf (C5)] There exists a positive number $L_{\psi}$ such that 
\begin{displaymath} 
\left\|\psi(x_1)-\psi(x_2)\right\| \geq L_{\psi}\left\|x_1-x_2\right\|,
\end{displaymath}  
for all $x_1,x_2 \in \mathbb R^{m}.$ 
\end{enumerate}  

The following assertion is based on the results in \cite{Akh8}.

\begin{theorem} \label{period-doubling_theorem}
If conditions $(C1)-(C5)$ hold and $\left|\varepsilon\right|$ is sufficiently small, then there exists a neighborhood ${\cal U}$ of the orbitally stable limit cycle of (\ref{base-system}) such that  solutions of (\ref{intro_replicator}) which start inside ${\cal U}$ behave chaotically around the limit cycle. That is, the solutions are sensitive and there are infinitely many unstable periodic solutions. 
\end{theorem}

\section{Economic Models: The Base Systems} \label{econ_sec3}

In what follows, we will require regular systems, that is, models with asymptotically stable equilibria or limit cycles, that can be perturbed to generate chaotic business cycles. In this part of the paper we propose three economic models to be used as base systems.

\subsection{Kaldor-Kalecki model with a steady equilibrium}

Consider the following model of an aggregate economy:
\begin{eqnarray}
\begin{array}{l}\label{kk}
Y'= \alpha [I(Y,K) - S(Y,K)], \\
K'= I(Y,K)  - \delta K,
\end{array}
\end{eqnarray}
\noindent where $Y$  is income, $K$ is capital stock, $I$ is gross investment, and $S$ is savings. Income changes proportionally to the excess demand in the goods market, and the second equation is a standard capital accumulation equation. The constant depreciation rate $\delta$ and the adjustment coefficient $\alpha$ are positive. This model was studied in detail in \cite{lorenz} and \cite{zhang}. It admits a stable equilibrium under certain conditions on the functions involved. 

Let us consider the following specification of system (\ref{kk}) with $I(Y,K) = Y - aY^3 + bK,$ $S(Y,K) = sY,$ 
\begin{eqnarray}
\begin{array}{l}\label{kk1}
 Y'= \alpha [(1-s)Y  - aY^3 + bK], \\
 K' = Y - aY^3  + bK - \delta K,
\end{array}
\end{eqnarray}  
where the constant parameters satisfy $a > 0,$ $b<0,$ $0 <s <1$ and $0 < \delta <1.$ 

One can see that a steady state of (\ref{kk1}) with  positive coordinates
\[Y^* = \sqrt{ \frac{\delta(1-s) + bs}{a \delta}}, \hspace{0.2in}K^* = \frac{s}{\delta} \sqrt{ \frac{\delta(1-s) + bs}{a \delta}},\]
exists only if  $\displaystyle \delta s<\delta+bs.$

The transformations $Y = y + Y^*,$ $K = k + K^*,$ applied to (\ref{kk1}), give us the system 
\begin{eqnarray}\label{KK_model}
\begin{array}{l}  
y'=\displaystyle \alpha \left[ \left( 2(s-1) - \frac{3bs}{\delta}  \right)y + bk -ay^3 -3\sqrt{\frac{a\delta(1-s)+abs}{\delta}}y^2    \right], \\
k'= \displaystyle \left( 3s-2-\frac{3bs}{\delta}   \right) y + (b-\delta) k -ay^3 -3\sqrt{\frac{a\delta(1-s)+abs}{\delta}}y^2.
\end{array}
\end{eqnarray}

\subsection{A model with a business cycle} 
 
We also investigate the idealized macroeconomic
model with foreign capital investment, 
\begin{eqnarray}
\begin{array}{l} \label{prib}
S'= \alpha Y+ pS(k-Y^2),\\
Y'=v(S+F),\\
F'=m S-r Y,
\end{array}
\end{eqnarray}
where $S(t)$  are savings of households, $Y (t)$ is Gross Domestic Product (GDP), $F(t)$ is
foreign capital inflow, $k$ is potential GDP, and $t$ is time.   If $k$  is   set to 1,   then  $Y,$ $S,$ $F$ are measured as multiples of potential output. The parameters represent corresponding ratios: $\alpha$ is the variation of the
marginal propensity to save, $p$ is the ratio of capitalised profit, $\displaystyle \frac{1}{v}$ is the capital-output ratio, $m$ is the capital inflow-savings ratio and $r$  is the debt refund-output ratio. The model in (\ref{prib}) was introduced by Bouali \cite{Bouali1}, and later studied by Bouali et al. \cite{Bouali2} and Pribylova \cite{Pribylova}.

Consider  system  $(\ref{prib})$  with  specified coefficients, 
\begin{eqnarray}
\begin{array}{l} \label{cycle2}
S'=\alpha Y+0.1S(1-Y^2),\\
Y'=0.5(S+F),\\
F'=0.19S-0.25Y.
\end{array}
\end{eqnarray}
According to \cite{Pribylova}, the system (\ref{cycle2}) admits Hopf bifurcation at $\alpha=\alpha_0 \equiv 0.25$ and an orbitally stable cycle appears as $\alpha$  decreases.

\subsection{A Kaldor-Kalecki model} 

Let us consider the system, 
\begin{eqnarray} \label{kkd11}
&& Y' = \alpha [ I(Y,K) - S(Y,k)], \nonumber\\
&& K' =  I(Y(t-\tau),K)-\delta K.
\end{eqnarray}
System (\ref{kkd11}) is a Kaldor  model with time delay. Kalecki \cite{Kalecki35} introduced the idea that there may be a time lag between the time an investment decision is made and the time investment is realised. The Kaldor-Kalecki model (\ref{kkd11}) was formalised by Krawiec and Szydlowski \cite{Krawiec99}, where investment depends on income at the time investment decisions are taken and on capital stock at the time investment is finished. One can find additional information on the models with delay in the papers \cite{Szydlowski01,Wang09}.

We will study the specification
\begin{eqnarray} \label{kkd}
&& Y' =  1.5[  \tanh(Y) -0.25 K - (4/3) Y], \nonumber\\
&& K' =  \tanh(Y(t-\tau)) -0.5 K.
\end{eqnarray}

According to Zhang and Wei \cite{Zhang04}, the model admits an orbitally stable limit cycle for $\tau >5.4.$ More precisely, the periodic solution appearance follows a Hopf bifurcation so that the origin is asymptotically stable if $\tau<5.4,$ and the origin loses its stability and the cycle bifurcates from the origin for $\tau > 5.4.$

\section{Extension of Chaos in a Constellation of Economical Models} \label{econ_sec4}

\begin{figure}[ht]
\centering
\includegraphics[height=2.0cm]{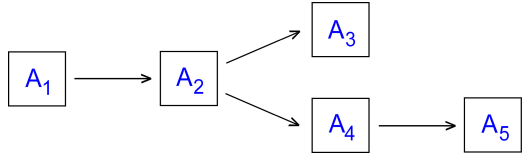} 
\caption{The connection topology of the systems $A_1-A_5.$}
\label{largesystem}
\end{figure}

To provide a comprehensive illustration for the discussion in the previous sections, we will consider a constellation of five unilaterally connected economic models denoted by $A_k,$ $k=1,\ldots,5.$ The topology of the connection is presented in Figure \ref{largesystem}, and the models are formulated in system (\ref{extension_example}). We will show that the chaos that appears in $A_1$ spreads to all the other models. $A_2$ serves as a replicator of the chaos of $A_1$ and as a generator of chaos in $A_3$ and $A_4.$ Model $A_4$ is a replicator of the chaos of $A_2$ and a generator of chaos in $A_5.$

The following is a system of five unidirectionally coupled models $A_1 - A_5.$
\begin{eqnarray}
\begin{array}{l} \label{extension_example}
\left.
\begin{aligned}
\kappa_{j+1}=\mu \kappa_j (1-\kappa_j), \\
\end{aligned}
\right\} A_1\\
\left.
\begin{aligned}
y'_1 = \displaystyle  (1/8) y_1 - (5/16) k_1 - a_1 y_1^3 - \frac{3\sqrt{a_1}}{2} y_1^2 + \nu_1(t,\theta), \\
k'_1 = \displaystyle (1/4) y_1 - (3/8) k_1 -  a_1 y_1^3 - \frac{3\sqrt{a_1}}{2} y_1^2, \\
\end{aligned}\right\} A_2\\
\left.
\begin{aligned}
y'_2 = \displaystyle (1/3) y_2 - k_2 - a_2y_2^3 - \frac{\sqrt{6a_2}}{2} y_2^2 + 0.6 y_1(t)  +  \nu_2(t,\zeta), \\
k'_2 = \displaystyle (1/2) y_2 - (5/4) k_2 - a_2y_2^3 - \frac{\sqrt{6a_2}}{2} y_2^2,\\ 
\end{aligned}
\right\} A_3\\
\left.
\begin{aligned}
S'= 0.23 Y+0.1S(1-Y^2),\\
Y'=0.5(S+F)+2(y_1(t)+0.5),\\
F'=0.19S-0.25Y, \\  
\end{aligned} 
\right\}  A_4\\
\left.
\begin{aligned}
y'_3=\displaystyle  (3/5)y_3-(4/5)k_3-a_3y_3^3-\frac{3\sqrt{a_3}}{\sqrt{10}} y_3^2+0.01 Y(t), \\
k'_3=\displaystyle  (7/10)y_3-(9/10)k_3-a_3y_3^3-\frac{3\sqrt{a_3}}{\sqrt{10}}y_3^2, 
\end{aligned}
\right\} A_5 \\ 
\end{array}
\end{eqnarray}
where $a_1,$ $a_2,$ $a_3$ are constants and the  piecewise constant functions $\nu_1(t,\theta)$  and $\nu_2(t,\zeta)$ are defined  as follows:
\begin{eqnarray} \label{relay1_econcycle}
\nu_1(t,\theta)=\left\{\begin{array}{ll} 0.019, ~\textrm{if}  & \theta_{2j}  < t  \leq \theta_{2j+1},  \\
                                         0.002, ~\textrm{if}  & \theta_{2j-1}  < t \leq \theta_{2j}, 
\end{array} \right. 
\end{eqnarray} 
and
\begin{eqnarray} \label{relay2_econcycle}
\nu_2(t,\zeta)=\left\{\begin{array}{ll} 0.0006, ~\textrm{if}  & \zeta_{2j} < t  \leq \zeta_{2j+1}, \\
                                        0.0017, ~\textrm{if}  & \zeta_{2j-1} < t \leq \zeta_{2j}, 
\end{array} \right..
\end{eqnarray}

The sequence $\theta=\left\{\theta_j\right\},$  $j\in \mathbb Z,$ of the discontinuity instants of the function (\ref{relay1_econcycle}) satisfies the relation $\theta_j=j+\kappa_j,$ where the sequence $\left\{\kappa_j\right\}$ is a solution of the logistic map $A_1$ with $\kappa_0 \in [0,1].$ The sequence $\zeta=\left\{\zeta_j\right\},$ $j\in \mathbb Z,$ of the discontinuity instants of (\ref{relay2_econcycle}) satisfies the relation $\zeta_j= 2\sqrt{2} j$ for each $j.$

Examples of shocks of the form (\ref{relay1_econcycle}) and (\ref{relay2_econcycle}) are natural disasters and extreme events in general, such as market crashes. They take a finite number of values (an earthquake either happens or not), but their timing is irregular  or   regular.

\subsection{Description of the models $A_1$ to $A_5$ }
  
Equation $A_1$ is the logistic map, which will be used as the main source of chaos in system (\ref{extension_example}). The interval $[0,1]$ is invariant under the iterations of the map for
the parameter values $\mu \in (0,4],$ and for $\mu= 3.8$ it is chaotic through period-doubling cascade \cite{Sch99}. The logistic map plays a very important role in many fields of science, particularly in economics. It can be used to describe economic variables. In \cite{Bala98} the logistic map emerges as the law of motion of the price of the non-numeraire good in a simple discrete-time model of an exchange economy with two goods under Walrasian tatonnement. Benhabib and Day \cite{Benhabib82} showed that a logistic map describes optimal consumption in a simple overlapping generations model with
a quadratic utility function, and Mitra and Sorger \cite{Mitra99} proved that the logistic map can be the optimal policy function of a regular dynamic optimisation problem, if and only if the discount factor does not exceed $1/16.$

The logistic map $A_1$ is the generator of chaos for the global system (\ref{extension_example}) and as we mentioned above, a generator can be not only with continuous dynamics, but also with discrete, and even hybrid, i.e., combining both continuous and discrete. In fact the whole model (\ref{extension_example}) is an example of a hybrid system. 
 
System $A_2$ describes the aggregate economy of Country $1.$ It is a perturbed Kaldor model ((\ref{KK_model}), obtained by setting $\alpha=1,$ $s=1/8,$ $\delta=1/16$ and $b=-5/16.$ In the
absence of the perturbation function $\nu_1(t,\theta)$, the model possesses an asymptotically stable equilibrium provided that the number $a_1$ is sufficiently small. One can verify that the associated linear system admits complex conjugate eigenvalues $(-1\pm i)/8.$ The function (\ref{relay1_econcycle}) describes a rainfall shock that impacts on the agricultural sector and through it on the total output. The higher value of $\nu_1$ implies normal rainfall, while the lower value is drought, which leads to lower agricultural production and slower output growth. 

Using the results of \cite{Akh5,Akh9,Akh7}, one can state that the chaoticity of the logistic map $A_1$ with $\mu= 3.8$ makes the function $\nu_1(t,\theta)$ behave chaotically, and system $A_2$ is chaotic through period-doubling cascade for the same value of the parameter $\mu$. That is, it admits infinitely many unstable periodic solutions and exhibits sensitivity. For each natural number $p,$ the system possesses an unstable periodic solution with period $2p.$ Next, in its own turn system $A_2$ is the generator for the systems $A_3$ and $A_4.$

System $A_3$ reflects the dynamics of Country $2.$ It is obtained by using the coefficients $\alpha=1,$ $s=1/6,$ $\delta=1/4,$ $b=-1$ in the Kaldor model (\ref{KK_model}) and by perturbing it with the solutions of $A_2$ as well as with the periodic function (\ref{relay2_econcycle}). The associated linear system has the eigenvalues $(-11\pm\sqrt{73})/24.$ In the absence of the perturbation
terms $0.6 y_1(t)$  and $\nu_2(t,\zeta)$ and if the number $a_2$ is sufficiently small, the system admits an asymptotically stable equilibrium. The term $0.6 y_1(t)$ describes the effect exports
from Country $2$ to Country $1,$ modelled as a function of the income of Country $1,$ $y_1(t),$ have on the rate of change in the income of Country $2.$ The function $\nu_2(t,\zeta)$ reflects productivity shocks in Country $2,$ which is a binary variable. The higher value of $\nu_2$ stands for faster productivity growth, and the lower value for slower productivity growth, which leads to slower output growth.

Since the periodic motions that are embedded in the chaotic attractor of system $A_2$ with $\mu = 3.8$ and the function (\ref{relay2_econcycle}) have incommensurate periods, one can confirm using the results of \cite{Akh6} that system $A_3$ is chaotic with infinitely many quasi-periodic solutions in the basis. This will be shown through simulations in Figure \ref{A3_zoom}.

System $A_4$ describes the aggregate economy of Country $3.$ It is obtained by perturbing system (\ref{cycle2}) with the solutions of $A_2.$ It is a replicator with respect to system $A_2,$ while the term $2(y_1(t)+0.5)$ is the input. This term represents the effect of exports from Country $3$ to Country $1,$ modelled as a function of income in Country $1,$ $y1(t),$ on the rate of growth of income in Country $3.$ 

In the absence of perturbations, $A_4$ possesses an orbitally stable limit cycle \cite{Pribylova}. Theorem \ref{period-doubling_theorem} implies that system $A_4$ admits chaotic business cycles, provided that the value of the parameter $\mu = 3.8$ is used in system $A_2.$ Since the orbitally stable cycle of system (\ref{cycle2}) occurs through a Hopf bifurcation, one can talk about the  \textit{bifurcation of the cyclic chaos}.

System $A_5$ models the dynamics of Country $4.$ It is constructed by perturbing the Kaldor model (\ref{KK_model}) with the solutions of $A_4,$ or in economic terms, by perturbing the
aggregate economy with exports from Country $4$ to Country $3,$ which are a fraction of the income of Country $3,$ $Y(t).$ The eigenvalues of the associated linear system are $-1/5$ and $-1/10.$ In the absence of the perturbation term $0.01 Y(t),$ the system possesses an asymptotically stable equilibrium, for sufficiently small values of $a_3.$ We will make use of system $A_5$ to demonstrate the \textit{attraction of chaotic business cycles}.

\subsection{Simulations}

In this part of the paper, we will demonstrate numerically the chaotic behavior in system (\ref{extension_example}). In what follows, we will use $a_1=3\times 10^{-6},$ $a_2=10^{-6},$ $a_3=5\times 10^{-6},$ $\mu=3.8$ and $\kappa_0=0.63.$

\begin{figure}[ht]
\centering
\includegraphics[height=3.2cm]{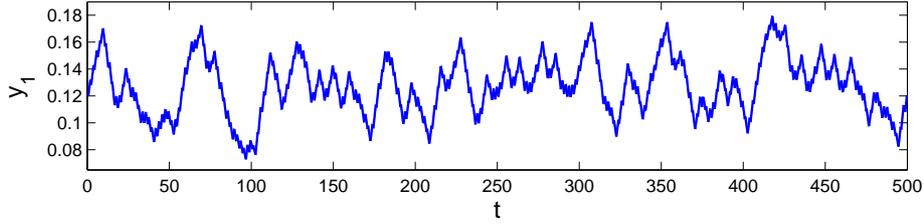} 
\caption{The graph of the $y_1$ coordinate of system $A_2.$}
\label{A2_withoutcontrol}
\end{figure}

Let us start with system $A_2.$ Setting the initial data $y_1(t_0)=0.12,$  $k_1(t_0)=0.08,$  where $t_0=0.63,$ we graph in Figure \ref{A2_withoutcontrol} the $y_1$ coordinate of system $A_2.$ It is seen in the figure that system $A_2$ behaves chaotically.

\begin{figure}[ht]
\centering
\includegraphics[height=3.2cm]{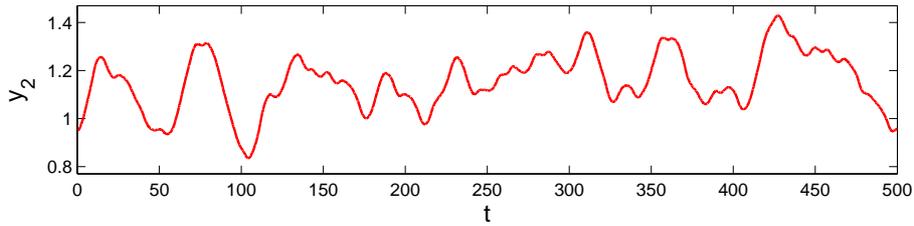} 
\caption{Extension of chaos by system $A_3.$}
\label{A3_withoutcontrol}
\end{figure}

\begin{figure}[ht]
\centering
\includegraphics[height=7.0cm]{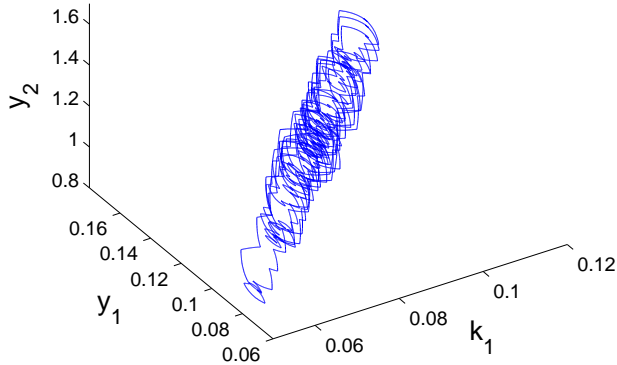} 
\caption{The projection of the chaotic trajectory of the coupled Kaldor-Kalecki system $A_2-A_3$ on the $y_1-k_1-y_2$ space.}
\label{chaotic_torus}
\end{figure}

To show the extension of chaos by system $A_3,$ we use the solution in Figure \ref{A2_withoutcontrol} as perturbation in system $A_3$ and present in Figure \ref{A3_withoutcontrol} the time series of the $y_2$ coordinate of $A_3$. The initial data $y_2(t_0)=0.95,$ $k_2(t_0)=0.38,$ where $t_0=0.63,$ is used in the simulation. Figure \ref{A3_withoutcontrol} reveals that the chaos of system $A_2$ is extended such that the system $A_3$ also possesses chaos. In order to confirm the extension of chaos once more, we depict in Figure \ref{chaotic_torus} the projection of the trajectory of the coupled Kaldor system $A_2-A_3,$ corresponding to the same initial data, on the $y_1-k_1-y_2$ space.

\begin{figure}[ht]
\centering
\includegraphics[height=6.0cm]{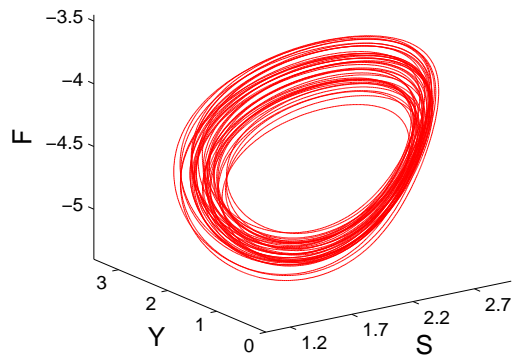} 
\caption{Chaotic business cycle of system $A_4.$}
\label{A4_cyclic}
\end{figure}

Next, we continue with system $A_4.$ We take into account system $A_4$ with the solution of $A_2$ that is represented in Figure \ref{A2_withoutcontrol}, and show the trajectory of $A_4$ with $S(t_0)=1.67,$ $Y(t_0)=0.94,$ $F(t_0)=-5.15,$ where $t_0=0.63,$ in Figure \ref{A4_cyclic}. One can observe in Figure \ref{A4_cyclic} that the system $A_4$ admits a chaotic business cycle.

\begin{figure}[ht]
\centering
\includegraphics[height=6.0cm]{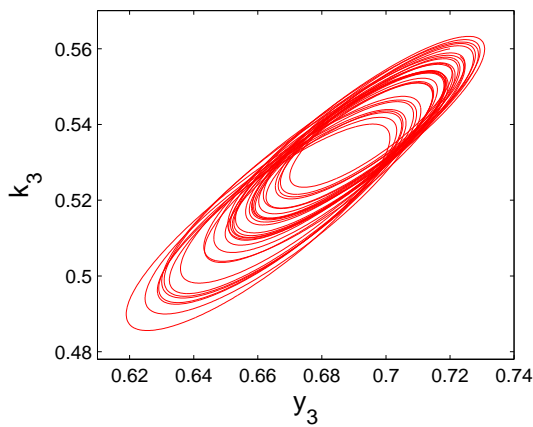} 
\caption{Attraction of cyclic chaos by system $A_5.$}
\label{A5_attraction}
\end{figure}

In order to observe the attraction of the cyclic chaos of system $A_4,$ we again use the solution of $A_4$ with $S(t_0)=1.67,$ $Y(t_0)=0.94,$ $F(t_0)=-5.15,$ where $t_0=0.63,$ in $A_5$ and depict in Figure \ref{A5_attraction} the trajectory of system $A_5$ with $y_3(t_0)=0.72,$ $k_3(t_0)=0.56.$ It is seen in Figure \ref{A5_attraction} that the chaotic business cycle of $A_4$ is attracted by $A_5,$ and the cyclic irregular behavior is extended.
 
\subsection{Control of extended chaos} 

The source of the chaotic motions in system (\ref{extension_example}) is the logistic map $A_1.$ Therefore, to control the chaos of the entire system, one has to stabilize an unstable periodic solution of the logistic map. The OGY control method \cite{Ott90} is one of the possible ways to do this. We proceed by briefly explaining the  method. 

Suppose that the parameter $\mu$ in the logistic map $A_1$ is allowed to vary in the range $[3.8-\epsilon, 3.8+\epsilon]$, where $\epsilon$ is a given small number. That is, it is not possible (say, it is prohibitively costly or practically infeasible) to simply shift the value of $\mu$ to a level that generates non-chaotic dynamics. Let us consider an arbitrary solution $\left\{\kappa_j\right\},$ $\kappa_0\in[0,1],$ of the map and denote by $\kappa^{(q)},$ $q=1,2,\ldots ,p,$ the target unstable $p-$periodic orbit to be stabilized. 
In the OGY control method \cite{Sch99}, at each iteration step $j$ after the control mechanism is switched on, we consider the logistic map with the parameter value $\mu=\bar \mu_j,$ where
\begin{eqnarray}\label{control_econcycle}
\bar \mu_j=3.8 \left[1+\frac{(2\kappa^{(q)}-1)(\kappa_{j}-\kappa^{(q)})}{\kappa^{(q)}(1-\kappa^{(q)})} \right],
\end{eqnarray}
provided that the number on the right-hand side of the formula $(\ref{control_econcycle})$ belongs to the interval $[3.8-\epsilon, 3.8+\epsilon].$ In other words, we apply a perturbation in the amount of $ \displaystyle\frac{3.8(2\kappa^{(q)}-1)(\kappa_{j}-\kappa^{(q)})}{\kappa^{(q)}(1-\kappa^{(q)})}$ to the parameter $\mu=3.8$ of the logistic map, if the trajectory $\left\{\kappa_j\right\}$ is sufficiently close to the target periodic orbit. This perturbation makes the map behave regularly so that at each iteration step the orbit $\kappa_j$ is forced to be located in a small neighborhood of a previously chosen periodic orbit $\kappa^{(q)}.$ Unless the parameter perturbation is applied, the orbit $\kappa_j$ moves away from $\kappa^{(q)}$ due to the instability. If $\displaystyle   \left|\frac{ 3.8(2\kappa^{(jq)}-1)(\kappa_{j}-\kappa^{(q)})}{\kappa^{(q)}(1-\kappa^{(q)})} \right| > \varepsilon$, we set $\bar \mu_{j}=3.8,$ so that the system evolves at its original parameter value, and  wait until the trajectory $\left\{\kappa_j\right\}$ enters a sufficiently small neighborhood of the periodic orbit $\kappa^{(q)},$ $q=1,2,\ldots ,p,$ such that the inequality $-\epsilon \le \displaystyle\frac{3.8 (2\kappa^{(q)}-1) (\kappa_{j}-\kappa^{(q)})}{\kappa^{(q)}(1-\kappa^{(q)})}  \le \epsilon$ holds. If this is the case, the control of chaos is not achieved immediately after switching on the control mechanism. Instead, there is a transition time before the desired periodic orbit is stabilized. The transition time increases if the number $\epsilon$ decreases \cite{Gon04}.

\begin{figure}[ht]
\centering
\includegraphics[height=3.2cm]{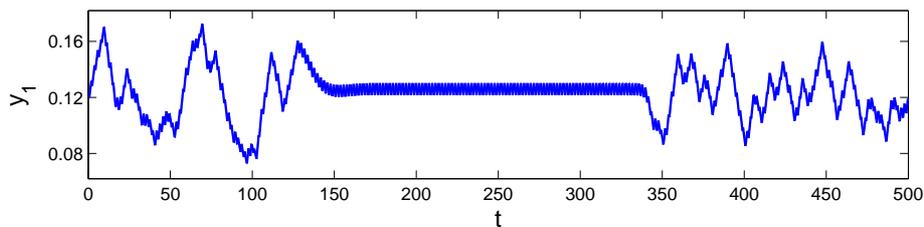} 
\caption{The chaos control of system $A_2.$ The OGY control method is applied around the fixed point $2.8/3.8$ of the logistic map. The value $\epsilon = 0.08$ is used.}
\label{A2_withcontrol}
\end{figure}

The chaos of system $A_2$ can be stabilized by controlling an unstable periodic orbit of the logistic map $A_1,$ since the map gives rise to the presence of chaos in the system. By applying the OGY control method around the fixed point $2.8/3.8$ of the logistic map, we stabilize the corresponding unstable $2-$periodic solution of system $A_2.$ The simulation result is seen in Figure \ref{A2_withcontrol}. We used the same initial data as in Figure \ref{A2_withoutcontrol}. It is seen in Figure \ref{A2_withcontrol} that the OGY control method successfully controls the chaos of system $A_2.$ The control is switched on at $t=\theta_{50}$ and switched off at $t=\theta_{280}.$ The values $\kappa_0=0.63$ and $\epsilon = 0.08$ are utilized in the simulation. The control becomes dominant approximately at $t=150$ and its effect lasts approximately until $t=340,$ after which the instability becomes dominant and irregular behavior develops again.

\begin{figure}[ht]
\centering
\includegraphics[height=3.2cm]{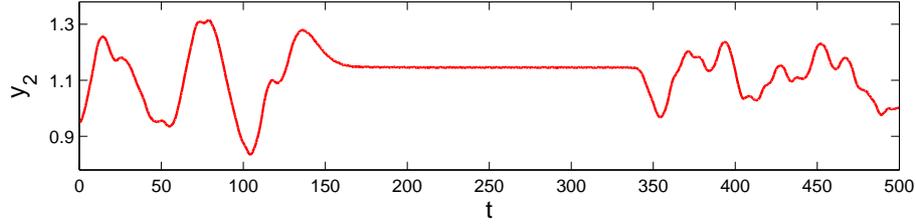} 
\caption{The chaos control of system $A_3.$ It is observable in the figure that controlling the chaos of system $A_3$ makes the chaos of system $A_2$ to be also controlled.}
\label{A3_withcontrol}
\end{figure}

Next, we will demonstrate the stabilization of an unstable quasi-periodic solution of system $A_3.$  We suppose that an unstable quasi-periodic solution of $A_3$ can be stabilized by controlling the chaos of system $A_2$. We use the solution shown in Figure \ref{A2_withcontrol} as the perturbation in system $A_3,$ and represent in Figure \ref{A3_withcontrol} the solution of $A_3$ with $y_2(t_0)=0.95,$ $k_2(t_0)=0.38,$ where $t_0=0.63.$ Similarly to system $A_2,$ it seen in the figure that the chaos of $A_3$ is controlled approximately for $150 \le t \le 340.$ 

\begin{figure}[ht]
\centering
\includegraphics[height=3.2cm]{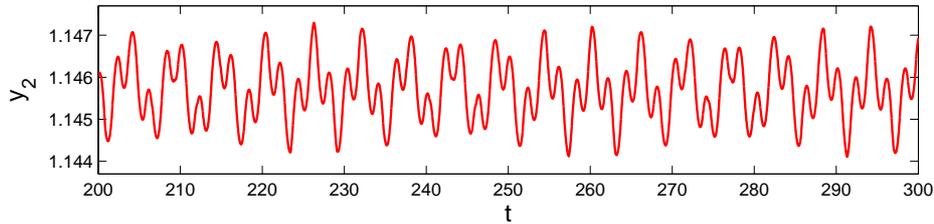} 
\caption{The stabilized quasi-periodic solution of system $A_3.$ }
\label{A3_zoom}
\end{figure}

To reveal that the stabilized solution is indeed quasi-periodic, we depict in Figure \ref{A3_zoom} the graph of the same solution for $200 \le t \le 300.$ Figure \ref{A3_zoom} manifests that application of the OGY control method to system $A_2$ makes an unstable quasi-periodic solution of $A_3$ to be stabilized. On the other hand, the stabilized torus of system $A_3$ is shown in Figure \ref{A3_zoom}.

\begin{figure}[ht]
\centering
\includegraphics[height=6.0cm]{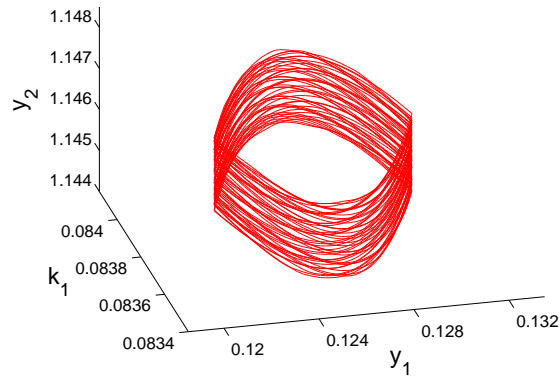} 
\caption{The stabilized torus of system $A_3.$}
\label{stabilized_torus}
\end{figure}

%%%%%%%%%%%%%%%%%%%%%%%%%%%%%%%%%%%%%%%%%%%%%%%%%%%%%%%%%%%%%%%%%%%%%%%%%%%

\section{Chaotic Business Cycles in Kaldor-Kalecki Model with Time Delay} \label{Kaldor_Kalecki_delay}
 
This section considers the phenomenon of chaos extension by utilizing an economical model with time lag (\ref{vdp_kk_delay}). We are devoting a separate discussion to this model, since the result for this case does not have theoretical support at the moment. The extension of chaos can be only observed numerically in our example, but in the future one could prove the entrainment of the limit cycle by chaos for functional differential equations using the results of the paper \cite{Akh8}. In this section, we will demonstrate numerically the formation of chaotic business cycles in the Kaldor-Kalecki model with time delay.
 
Let us take into account the system, 
\begin{eqnarray}
\begin{array}{l} \label{vdp_kk_delay}
\left.
\begin{aligned}
x''+5(x^2-1)x'+x=5\displaystyle\cos(2.467t),\\
\end{aligned}
\right\} B_1\\
\left.
\begin{aligned}
Y' =  1.5 \left[  \tanh(Y) -0.25 K - (4/3) Y   \right] +  0.0045 x(t), \\
K' =  \tanh(Y(t-\tau)) -0.5 K.  \\
\end{aligned}\right\} B_2\\
  \end{array}
\end{eqnarray}

Equation $B_1$ is the chaotic Van der Pol oscillator, which is used as the generator system in (\ref{vdp_kk_delay}). Van der Pol type equations have played a role in economic modelling \cite{Goodwin51,Goodwin90,lorenz}. It is shown by Parlitz and Lauterborn \cite{Parlitz87} that equation $B_1$ is chaotic through period-doubling cascade. The process of period-doubling is described by Thompson and Stewart \cite{Th02}. This implies that there are infinitely many \textit{unstable} periodic solutions of $B_1,$ all with different periods. Due to the absence of stability, any solution that starts near the periodic motions behaves \textit{irregularly}. We will interpret the solution $x(t)$ as an irregular productivity shock.

System $B_2$ is the Kaldor-Kalecki model and it is the result of the perturbation of the model (\ref{kkd}) of an aggregate economy with a productivity shock. We will observe numerically the appearance of a chaotic business cycle, and in particular, the entrainment by chaos of the limit cycle of system (\ref{kkd}), in the next simulations.

\begin{figure}[ht]
\centering
\includegraphics[height=5.5cm]{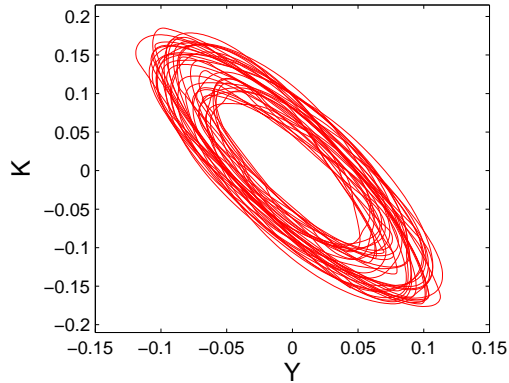} 
\caption{The appearance of chaotic business cycle in the Kaldor-Kalecki model $B_2.$}
\label{cyclic_kaldor_kalecki_delay}
\end{figure}

Let us take $\tau=5.5$ in $B_2$ so that the system possesses an orbitally stable limit cycle in the absence of perturbation \cite{Zhang04}. We make use of the solution $x(t)$ of $B_1$ with $x(0)=1.1008,$ $x'(0)=-1.5546,$ and present in Figure \ref{cyclic_kaldor_kalecki_delay} the solution of $B_2$ with the initial condition $Y(t)=u(t)$ and $K(t)=v(t)$ for $t\in[-\tau,0],$ where $u(t)=-0.057$ and $v(t)=0.063$ are constant functions. Figure \ref{cyclic_kaldor_kalecki_delay} reveals that the dynamics of $B_2$ exhibits chaotic business cycles. This result shows that our theory of chaotic business cycles can be extended to systems with time delay.

%%%%%%%%%%%%%%%%%%%%%%%%%%%%%%%%%%%%%%%%%%%%%%%%%%%%%%%%%%%%%%%%%%%%%%%%%%%

\section{Chaos Extension Versus Synchronization} \label{econ_sec6}

Generalized synchronization characterizes the dynamics of a response system that is driven by the output of a chaotic driving system \cite{Abarbanel96,Gon04,Hunt97,Kocarev96,Rulkov95}. Suppose that the dynamics of the drive and response are governed by the following systems with a skew product structure
\begin{eqnarray} \label{syncdrive}
x'=D(x)
\end{eqnarray}
and
\begin{eqnarray} \label{syncresponse}
y'=R(y,K(x)),
\end{eqnarray}
respectively, where $x\in\mathbb R^p,$ $y\in\mathbb R^q.$ Synchronization \cite{Rulkov95} is said to occur if there exist sets $I_x,$ $I_y$ of initial conditions and a transformation $\phi,$ defined on the chaotic attractor of (\ref{syncdrive}), such that for all $x(0)\in I_x,$ $y(0)\in I_y$ the relation $\displaystyle \lim_{t\to\infty} \left\|y(t)-\phi(x(t))\right\|=0$ holds. In this case, a motion that starts on $I_x\times I_y$ collapses onto a manifold $M\subset I_x\times I_y$ of synchronized motions. The transformation $\phi$ is not required to exist for the transient trajectories. When $\phi$ is the identity, the identical synchronization takes place \cite{Pecora90,Gon04}.

It is formulated by \cite{Kocarev96} that generalized synchronization occurs if and only if for all $x_0\in I_x,$ $y_{10},y_{20}\in I_y,$ the following asymptotic stability criterion holds:
$$
\displaystyle \lim_{t\to\infty} \left\| y(t,x_0,y_{10}) - y(t,x_0,y_{20})  \right\|=0,
$$
where $y(t,x_0,y_{10}),$ $y(t,x_0,y_{20})$ denote the solutions of (\ref{syncresponse}) with $y(0,x_0,y_{10})=y_{10},$ $y(0,x_0,y_{20})=y_{20}$ and the same $x(t),$ $x(0)=x_0.$

A numerical method that can be used to investigate coupled systems for generalized synchronization is the auxiliary system approach \cite{Abarbanel96,Gon04}. Let us investigate the coupled economic model $A_2-A_4$ for generalized synchronization by means of the auxiliary system approach. 

Consider the auxiliary system
\begin{eqnarray} \label{auxiliary_Prybilova}
\begin{array}{l}
S_0'= 0.23 Y_0+0.1S_0(1-Y_0^2),\\
Y_0'=0.5(S_0+F_0)+2y_3(t),\\
F_0'=0.19S_0-0.25Y_0.
\end{array}
\end{eqnarray} 
System (\ref{auxiliary_Prybilova}) is an identical copy of system $A_4.$

\begin{figure}[ht]
\centering
\includegraphics[height=5.5cm]{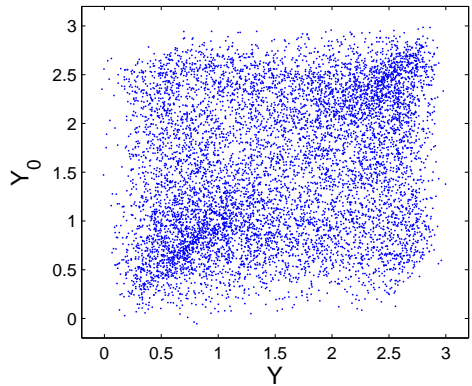} 
\caption{The auxiliary system approach shows that the systems $A_2$ and $A_4$ are not synchronized in the generalized sense.}
\label{aux_system_Prybilova}
\end{figure}

By marking the trajectory of system $A_2-A_4-(\ref{auxiliary_Prybilova})$ with initial data $y_1(t_0)=0.12,$ $k_1(t_0)=0.08,$  $S(t_0)=1.67,$ $Y(t_0)=0.94,$ $F(t_0)=-5.15,$ $S_0(t_0)=2.63,$ $Y_0(t_0)=0.84,$ $F_0(t_0)=-2.89$ at times $t=\theta_j$ and omitting the first $500$ iterations, we obtain the stroboscopic plot whose projection on the $Y-Y_0$ plane is shown in Figure \ref{aux_system_Prybilova}. Since the plot is not placed on the line $Y_0 = Y,$ we conclude that generalized synchronization does not occur in the couple $A_2-A_4.$

It is worth noting that generalized synchronization does not take place also in the dynamics of the unidirectionally coupled subsystems $B_1$ and $B_2,$ which are mentioned in Section  \ref{Kaldor_Kalecki_delay}, and this can be verified by means of the auxiliary system approach \cite{Abarbanel96,Gon04} as well.

%%%%%%%%%%%%%%%%%%%%%%%%%%%%%%%%%%%%%%%% 
 
\section{The Global Unpredictability, Synergetics  and Self-Organization} \label{econ_sec7}

The idea of the transition of chaos from one system to another, as well as the arrangement of chaos in an ordered way, can be viewed through the lens of self-organization \cite{haken83,nicolis}. Durrenmatt \cite{durrenmatt} explained that ``... a system is self-organizing if it acquires a spatial, temporal or functional structure without specific interference from the outside. By `specific' we mean that the structure of functioning is not impressed on the system, but the system is acted upon from the outside in a nonspecific fashion." There are three approaches to self-organization, namely thermodynamic (dissipative structures), synergetic and the autowaves approach. For the theory of dynamical systems (e.g. differential equations) the phenomenon means that an autonomous system of equations admits a regular and stable motion (periodic, quasiperiodic, almost periodic). These are what are called autowaves processes \cite{avk} or self-excited oscillations \cite{Moon} in the literature. We are inclined to add to the list one more phenomenon - chaos extension. For example, consider the collection of systems $A_1, A_2, \ldots, A_5$ once again, where $A_1$ is the original generator of chaos. Because of the connections and the conditions discovered in our analysis, all the other subsystems, $A_2, \ldots, A_5,$  are also chaotic. We believe this is a self-organization phenomenon, that is, a coherent behavior of a large number of systems \cite{haken83}.

Haken \cite{haken83}, a German theoretical physicist, introduced a new interdisciplinary field of science, synergetics, which deals with the origins and the evolution of spatiotemporal structures. Synergetics is based in large part on the dynamical systems theory. One of the crucial features of systems in synergetics is self-organization, which was discussed above. According to Haken \cite{haken83}, the central question in synergetics is whether there are general principles which govern the self-organized formation of structures and/or functions. The main concepts of the theory are instability, order
parameters, and slaving \cite{haken83}.

Instability is understood as the formation or collapse of structures (patterns) \cite{nicolis}. This is very common in fluid dynamics, lasers, chemistry and biology \cite{haken83,murray,nicolis,vorontsov}. A number of examples of instability can be found in the literature on morphogenesis \cite{Turing}, and pattern formation examples can be found in fluid dynamics. The phenomenon is called instability because a former state of fluid transforms into a new one, loses its ability to persist, and becomes unstable. One can view the formation of chaos in systems $A_2,\ldots, A_5$ in our results as instability. Even though processes in finite dimensional spaces are considered, chaotic attractors are assumed to be not single trajectories, but collections of infinitely many trajectories with complex topologies. One might say that they are somehow in-between objects of ordinary differential equations and partial differential equations. This allows us to also talk about dissipative structures \cite{nicolis}, due to the ``density" of the chaotic trajectories in the space.

Order parameters, when applied to differential equations theory, are those phase variables whose behavior produces the main properties of a macroscopic structure and which dominate all other variables in the formation, so that the latter can even depend on the order parameters functionally. The dependence that is proved (discovered) mathematically is what is called slaving. It is not difficult to see that the variables of system $A_1$ are order parameters, and they determine the chaotic behavior of the joined systems' variables.

\section{Conclusion} \label{econ_sec8}

We provide examples of models of aggregate economy where the main variables exhibit cycle-like motion with chaotic elements. Thus, we obtain an irregular business cycle in a deterministic setting. This provides a modelling alternative to the business cycle literature relying on stochastic variation in the economy. Additionally, our investigation highlights the variety of ways of generating chaos in an economic model. Previous work has focused on generating chaos and, in particular, chaotic business cycles \textit{endogenously} (see \cite{Bouali1,Bouali2,lorenz,Rosser,zhang}). Our method of creating chaos has its own relevance for economics, since we show the role of \textit{exogenous} shocks in the appearance of chaos in models that otherwise do not exhibit irregular behavior. It can also be said that our work provides a missing link in the research on the origins of irregularities in economic time series. While the literature on endogenous chaos was a response to the view that exogenous stochastic
shocks are the source of fluctuations in the economy (see \cite{Baumol}), this paper is a response to the former, in that it provides a role for exogenous chaotic disturbances in producing these fluctuations, and thus completes the circle.

Baumol and Benhabib \cite{Baumol} summarized the significance of chaos research for economics: ``Chaos theory   has  at  least  equal  power  in providing  caveats for  both  the economic analysis  and the policy  designer. For example,  it  warns us that  apparently  random behavior  may  not  be random at  all.   It  demonstrates   dramatically the dangers of extrapolation and the  difficulties  that  can beset economic forecasting generally.  It  provides the basis  for  the construction of simple models  of the behavior of rational  agents,   showing how even  these can yild extremely   complex developments. It  has served  as the basis for  models of learning  behavior  and has been  shown  to  arise naturally in a number of standard equilibrium models. It  offers additional insights about  the  economic source of oscillations in  a number  of economic models.''

Indeed, applications of chaos theory have illustrated the possibility of producing complex dynamics in deterministic settings \cite{Bouali1, Bouali2,Gabisch,Goodwin90,lorenz,Pribylova,zhang}, with some papers specifically focusing on building ``chaotic business cycles" \cite{Fanti}. Chaos is generated endogenously, and its appearance hinges on the values of some crucial parameters of the model. The main novelty of our paper is that we start with a model that is not endogenously complex. In one case (model $A_4$ in the main body), we assume that the system has a limit cycle, where the limit cycle is understood to be a closed orbit that is also an attractor \cite{Hirsch}. We then subject the model to chaotic exogenous shocks and obtain a perturbed system that admits chaotic motions. The chaos emerging around the original limit cycle is cycle-like, and therefore can be called a chaotic business cycle. This approach is based on rigorous mathematical theory \cite{Akh5,Akh7}, and we provide numerical
simulations. In another case (model $A_5$ in the main body), we subject a system with an asymptotically stable equilibrium to chaotic cyclic shocks, which produces a chaotic business cycle in the original model, as well. We demonstrate this scenario with simulations, as this approach does not have theoretical underpinnings as yet.

In this paper we show that it is possible to produce a chaotic business cycle in a very natural way - take a system of differential equations with a limit cycle as a point of departure, and introduce a chaotic exogenous disturbance. An example of an exogenous disturbance is a technology shock to the economy which affects output, holding all other variables constant. We describe it using  solutions of   chaos generator  models. We use them  to demonstrate the proposed approach, and other formulations can be studied in future work. For example, one can use actual economic time series, such as commodity prices, that have been tested for deterministic chaos \cite{Barnett, Brock,Frank,Panas}. Moreover, shocks other than technology shocks can be considered, in view of the on-going debate between two literatures supporting and rejecting the importance of technology shocks for generating business cycles \cite{Atella,Gali, Kydland}. 

Our results give more theoretical lights on the processes, as we suggest a mathematical apparatus, which describe rigorously \textit{extension of chaos}, increases its complexity, and  provides new structures of \textit{effective control} for clusters of economic models.

\vspace{0.5cm}
\noindent{\bf{Acknowledgements}} 

\vspace{0.35cm}

Z. Akhmetova is supported by a grant from the School of Economics, ASB, UNSW, Sydney, Australia. M.O. Fen is supported by the 2219 scholarship programme of T\"{U}B\.{I}TAK, the Scientific and Technological Research Council of Turkey.

\end{document}